	\documentclass[apj]{emulateapj}

	\usepackage{apjfonts}
	\usepackage{natbib}
	\usepackage{graphicx}
	\usepackage{amssymb}
	\usepackage{amsmath}
	
	\bibpunct{(}{)}{;}{a}{}{,}
\begin{document}
	\title{Imaging Spectropolarimetry with IBIS II:\\
	on the fine structure of G-band bright features}
	\shorttitle{On the fine structure of G-band bright features}
		
	\author{B.~Viticchi\'{e}\altaffilmark{1,2},
		D.~Del~Moro\altaffilmark{2},
		S.~Criscuoli\altaffilmark{3},
		F.~Berrilli\altaffilmark{2}}
	\shortauthors{Viticchi\'{e} et al.}
	
	\altaffiltext{1}{ESA/ESTEC RSSD, Keplerlaan 1, 2200 AG Noordwijk, Netherlands}
	\altaffiltext{2}{Dipartimento di Fisica, Universit\`a degli Studi di Roma
		``Tor Vergata'', Via della Ricerca Scientifica 1, I-00133 Rome, 
		Italy}
	\altaffiltext{3}{INAF-Osservatorio Astronomico di Roma, Via Frascati 33, I-00040,
		Monte Porzio Catone, Italia}

	\email{Bartolomeo.Viticchie@esa.int}

%---- Abstract
\begin{abstract}
	We present new results from first observations of the quiet solar
	photosphere performed through the Interferometric BIdimensional Spectrometer (IBIS)	
	in spectropolarimetric mode. 
	IBIS allowed us to measure the four Stokes parameters in the
	\ion{Fe}{1}~630.15~nm and \ion{Fe}{1}~630.25~nm lines
	with high spatial and
	spectral resolutions for $53$ minutes; the polarimetric sensitivity
	achieved by the instrument is $3\times10^{-3}$ the continuum intensity level.
	
	We focus on the correlation which emerges between
	G-band bright feature brightness and magnetic filling factor of $\sim10^3$~G (kG) fields
	derived by inverting Stokes $I$ and $V$ profiles.
	More in detail, we present the correlation first in a pixel-by-pixel study of
	a $\simeq3$~arcsec wide bright feature (a small network patch) and then
	we show that such a result
	can be extended to all the bright features found in the dataset
	at any instant of the time sequence.
	The higher the kG filling factor associated to a feature the higher the
	brightness of the feature itself. Filling factors up to
	$\simeq35$\% are obtained for the brightest features.
	Considering the values of the filling factors
	derived from the inversion analysis of spectropolarimetric data and
	the brightness variation observed in G-band data we put forward
	an upper limit for the smallest scale over which magnetic flux concentrations
	in intergranular lanes produce a G-band brightness enhancement
	($\simeq0.1''$).
	Moreover, the brightness saturation observed for feature
	sizes comparable to the resolution of the observations
	is compatible with large G-band bright features being
	clusters of sub-arcsecond bright points. This conclusion
	deserves to be confirmed by forthcoming spectropolarimetric
	observations at higher spatial resolution.
\end{abstract}

%---- Keywords
	\maketitle

%---- Paper
\section{Introduction}
\label{Intro}
	Solar magnetic fields manifest in the photosphere in a
	large variety of structures. At sub-arcsecond spatial
	scales they can appear concentrated in bright and roundish
	features, located in convective downflow regions, preferentially
	at the vertexes of granules. These bright features, first observed by
	\citet{DunZir73} and \citet{Meh74}
	in the network, are ubiquitous on the solar photosphere
	\citep{SanA04,deW05,BovWie08,SanA10}
	and their aggregation in active regions form, on larger spatial scales,
	faculae \citep[e.g.,][]{Fraz1972,Meh74,Rabin1992}. Their photometric properties
	are largely dependent on the spectral range in which they are observed
	\citep[e.g.,][]{Sut99,Lan2004,Trit2006, Bec07}. In particular,
	they show enhanced contrast with respect to the quiet Sun average
	intensity (typically $30$\%) when observed in the G-band (around $430.8$~nm)
	and are therefore referred to as G-band bright points
	\citep[][]{MulRoud1984SoPh}. 
	Due to their small size, the investigation of these features has been long hampered by
	spatial resolution limits imposed by instrumentation and atmospheric seeing.
	During the last decade, the development of Adaptive Optic Systems as well as
	post-facto restoration techniques \citep{Lof98,Lof02,VanN05}  has allowed
	studies that focused on the photometric
	\citep[e.g.,][]{Kel92,Ber95,BerTit96,BerTit01,Ber04,BovWie03,Ber07, Utz09, Kob09},
	dynamic  \citep[e.g.,][]{BovWie03, Nis03, deW05, Ish07, Lang07, deW08, Utz09b}
	and magnetic \citep[e.g.,][]{Mul00, BerTit01, Ber07, Ish07, Bha06, Bec07, deW08, Vit09}
	properties of these small-size magnetic flux concentrations. 
	In particular, observations have revealed that these
	are highly structured. \citet{Ber04} reported
	that in active regions, at locations of high magnetic flux density correspond
	in both Ca and G-band images a large variety  of features that they describe
	as filamentary crinkles-like structures, roundish flower- and ribbon-like
	features and, at the smallest spatial scales, point-like brightenings.
	Photometric properties and magnetic flux density are not constant within most
	of them, both contrast and magnetic flux density being in general higher at
	their edges and lower at their centers. The opposite is observed for
	point-like features. All these structures are highly dynamic and can
	evolve one into another \citep{Roup05,Vit09}. In quiet Sun regions,
	most of the features are point-like and are often clustered in
	groups or ``patches'' at the edges of granules and mesogranules \citep{SanA04,deW05,deW08}.
	The properties of G-band bright features (e.g., dimension and contrast)
	strongly depend on the angular resolution of observations so that
	it is hard to rigorously define the ``real bright point''.
	For this reason, in the following we will use the name ``bright feature''
	to indicate a local excess of brightening in the intergranular lanes of G-band filtergrams,
	independently of the properties of the feature itself.
	
	Theoretically, bright features were investigated through the
	model of flux-tube \citep{Spr1976}, a small-size concentration
	of magnetic field surrounded by field-free plasma. Due to
	the high magnetic pressure inside the kG flux tube, the gas pressure
	(gas density) is strongly reduced causing the photons
	to escape from deep photospheric layers that are hotter
	than the surrounding plasma in cold intergranular lanes.
	Numerical static and dynamic models of isolated
	\citep[][]{Dein1984, Knol88,Pizzo1993,Ste01,Ste05}
	or clustered \citep[][]{Cac79,OkKn05,CrisRas09} flux tubes, have
	proved to be successful in reproducing many of the observed
	properties, such as the enhancement of contrast toward the
	limb and in some molecular bands \cite[in particular CH and CN bands, e.g.,][]{Sch03},
	the presence of dark lanes which often accompany the bright
	features in off-disk center observations, and the suppression
	of convective motions within them.
	
	From the above figure follows that G-band bright features
	are proxies for kG fields in the solar photosphere.
	On the other hand, the inspection of movies reveals that
	the brightness of the features is not constant
	with time but intermittently can vary such that features can appear
	dark or bright at different instant. This evidence questions the
	flux tube evacuation as the principal physical mechanism responsible
	for contrast enhancement in small size magnetic concentrations.
	
	The understanding of the relation between the contrast
	measured in molecular bands and the properties of magnetic concentrations
	is of relevance for all those studies which employ broad-band
	imaging as proxy of strong magnetic fields \citep{deW05}. In particular
	G-band imaging of the solar photosphere, thanks to the high
	spatial resolution that can be achieved in these data, is an
	independent and complementary tool to spectropolarimetric observations.
	This is of great relevance, for instance, for the
	investigation of the presence of kG fields in the quiet Sun,
	which is still a debated topic in solar physics \citep[see][for a review]{deW09}.
	
	In \citet{Vit09} the first results from observations
	performed with IBIS  \citep[][]{Cav06} in spectropolarimetric
	mode revealed the advantages of combining high spatial resolution
	spectropolarimetry simultaneous and co-spatial to G-band imaging in
	the study of the temporal evolution of bright features in
	the quiet Sun. In the present work we report further results
	obtained from the same analysis concerning the relation between
	magnetic filling factor and brightness of G-band bright features.
	More in detail, these complete and
	corroborate the picture outlined in \citet{Vit09} only for three
	peculiar cases. The aim of this paper is
	to go more into the details of the capabilities of IBIS and
	to confirm the finding of the relation between kG magnetic filling
	factor and G-band bright feature intensity by exploiting the whole
	analyzed dataset. Such a relation will eventually allow us
	to conclude on the fine structuring of G-band bright features.
	
	In \S~\ref{Data} the dataset
	is presented, in \S~\ref{Methods} the adopted analysis methods
	are described, in \S~\ref{Results} we report the results from the analysis
	that are then discussed in \S~\ref{Disc}; in \S~\ref{Conc} the
	main conclusions are summarized.
	\begin{figure}[!ht]
	\centering
  	\includegraphics[width=7cm]{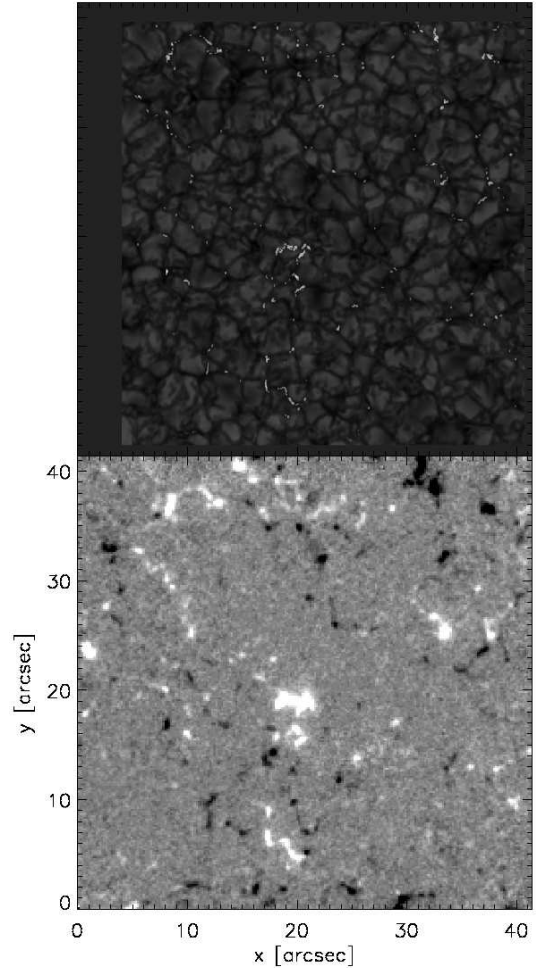}
 	\caption{Two snapshots from the analyzed dataset. \textit{Upper panel}:
 		G-band snapshot; bright feature intensity has been enhanced to show the result
 		of the automatic identification procedure. \textit{Lower panel}:
 		COG magnetogram saturated
 		at $\pm200$~G. The images refer to 16:42:29~UT.\label{fig1}}
 	\end{figure}
	
\section{Data}
\label{Data}
	A quiet region at disk center was observed on November 21, 
	2006 from 16:24~UT to 17:17~UT
	with IBIS, at the NSO/Dunn Solar Telescope. 
	The field-of-view (FOV)
	was approximatively $40''\times40''$.
	The acquired dataset consists of $50$ scans of the two \ion{Fe}{1}~$630$~nm
	lines performed with a cadence of $89$ seconds.
	The seeing during the acquisition of the first 
	$36$ scans (which have been analyzed in this work) was excellent and
	stable allowing the Adaptive Optics system 
	\citep{Rim04} to achieve near diffraction-limited performance.	
	The lines were sampled at $45$ wavelength points with a spectral FWHM of $2$~pm
	and a step of $2.3$~pm. In the scanning procedure the
	telluric line in between the two \ion{Fe}{1} lines	
	was skipped while the O$_2$~630.28~nm one was sampled with three
	points; this line allows us to set the absolute
	wavelength scale of our spectropolarimetry measurements.
	
	To perform spectropolarimetric measurements, the incoming light to IBIS
	is modulated by a pair of nematic liquid crystal variable retarders 
	placed in a collimated beam in front of
	the field stop of the instrument.
	The light is analyzed by a beam splitter in front of the
	detector, imaging two orthogonal states onto the same chip
	thus allowing for dual-beam spectropolarimetry.
	The modulation is in such a way that at each wavelength position
	six modulation states $I+S$ (and its orthogonal states $I-S$)
	are detected with the following temporal scheme: $S=[+V,-V,+Q,-Q,+U,-U]$.
	Spectropolarimetric images have a pixel scale of $0.18''$.
	Simultaneous to narrow-band spectropolarimetric data,
	broad-band ($633.32\pm5$~nm) counterparts of the same FOV were acquired.
	The exposure time for both narrow-band and broad-band data
	is $80$~ms.  The pixel scale of broad band data is $0.09''$.
	These reference images are normally used to
	register their spectropolarimetric counterparts.
	In this particular observation setup, a post processing
	Multi-Frame Blind Deconvolution procedure \citep[MFBD; ][]{Lof02}
	has been applied to these broad-band images to obtain a
	master frame from each scan with much reduced 
	seeing degradation and a homogeneous resolution in the whole FOV.
	We are then able to correct for anisoplanatism applying a de-stretch process on the single
	broad-band images, using this master frame as reference image.
	The computed de-stretch matrixes are successively applied also to the spectropolarimetric images.
	\begin{figure*}[!ht]
 	\centering
 	\includegraphics[height=5cm]{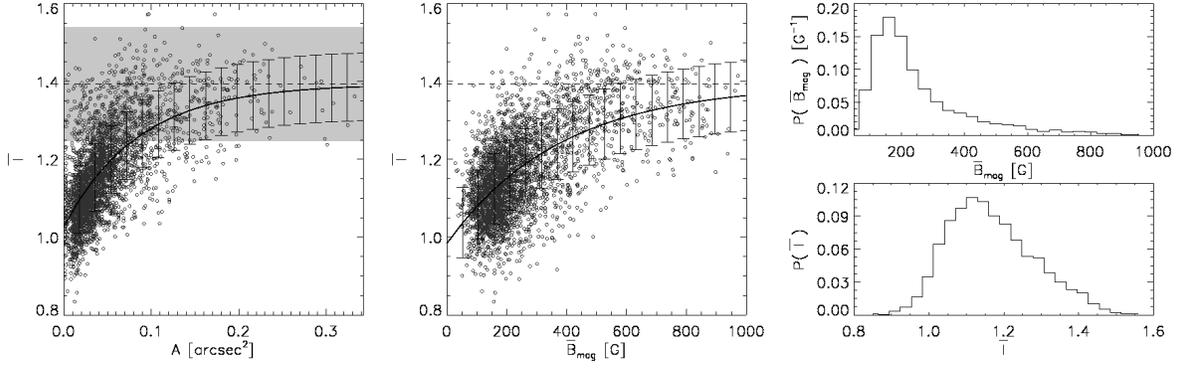}
  	\caption{Average properties of bright features from both G-band images and COG magnetograms.
  		The plots are derived by calculating: the average flux density $\bar{B}_{mag}$,
  		the average G-band intensity $\bar{I}$ (normalized to the average intensity of the time
  		sequence), and the total area $A$ of bright features
  		identified through the automatic procedure described in \S~\ref{Methods}.
		In the left panel: $\bar{I}$ vs. $A$ scatter plot.  		
  		In the central panel: $\bar{I}$ vs. $\bar{B}_{mag}$ scatter plot.
  		In both the plots an exponential fit of the data is reported (solid line).
		The error bars represent the average standard deviation error of the fit;
		the horizontal dashed line represent the saturation value $\bar{I}_{sat}$;
		the shaded area in the left panel represent the standard
		deviation error of $\bar{I}_{sat}$. For details on the fit procedure refer
		to \S~\ref{ResultsGbMag} and Table~\ref{tabfit}. In the right panels:
		histogram for $\bar{I}$ (upper panel) and $\bar{B}_{mag}$ (lower panel), respectively.
		\label{fig2}}
 	\end{figure*}
 	
 	\begin{figure}[!ht]
 	\centering
 	\includegraphics[height=17cm]{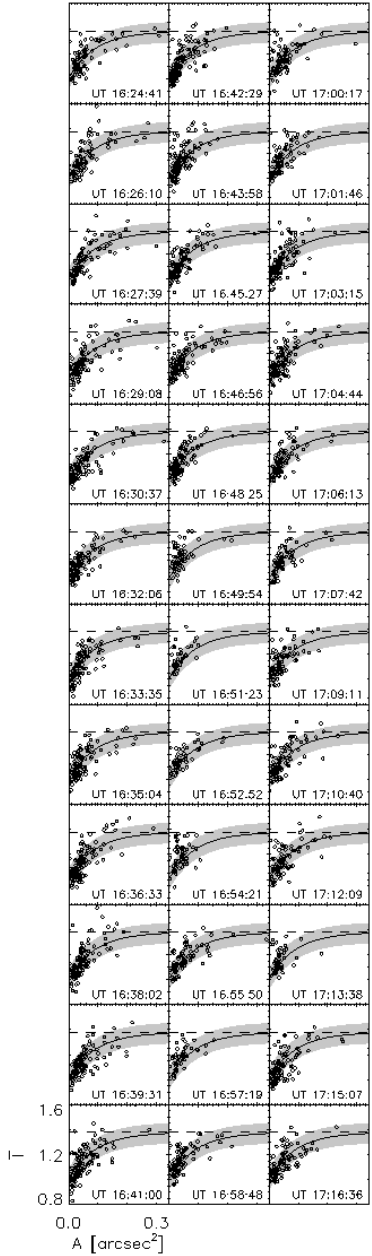}
 	\caption{$\bar{I}$ vs. $A$ scatter plots along the whole acquired time sequence;
 		$\bar{I}$ is normalized to the average G-band intensity of each frame.
 		The time sequence is divided in three columns
 		representing three subsequent time intervals; in each column the time rolls from up to down.
 		Each scatter plot relates $\bar{I}$ and $A$ for each instant of the time sequence.
 		In all the plots the result of the exponential fit shown in the left panel
 		of Fig.~\ref{fig2} is reported (solid line). The shaded area here represents the
 		average standard deviation with respect to the exponential fit (error bars in
 		Fig.~\ref{fig2}). The horizontal dashed line represents the saturation value $\bar{I}_{sat}$.
 		\label{fig7}}
 	\end{figure}
	\begin{deluxetable*}{lccc}
	\tabletypesize{\small}
	\tablecolumns{4} 
	\tablewidth{10cm} 
	\tablecaption{results from the exponential fits\label{tabfit}}
	\tablehead{	\colhead{scatter plot}	&
          		\colhead{$a$}	      	&
           		\colhead{$b$}	    	&
          		\colhead{$c$}      		}
    \startdata
	$\bar{I}$ vs. $A$				&$-0.37\pm0.13$ 	&$-14\pm7$~[arcsec$^{-2}$]			&$1.4\pm0.1$	\\
	$\bar{I}$ vs. $\bar{B}_{mag}$	&$-0.40\pm0.07$ 	&$-3.0\pm0.9$~[$10^{-3}$~G$^{-1}$] 	&-				\\
	$\bar{I}$ vs. $\bar{f}$			&$-0.37\pm0.04$ 	&$-7.0\pm0.9$						&-	
	\enddata
	\end{deluxetable*}
	
	Due to the classical mounting of the double etalon system,
	the spectral bandpass of IBIS experiences a wavelength-dependent blueshift within the FOV.
	For each pixel in the FOV, the blueshift amount
	is computed fitting a template line profile on
	the calibration dataset of flat-fields acquired just after the observations,
	taking into consideration also the
	displacements introduced by the de-stretch procedure.
	The calibration data includes flats, darks, resolution targets,
	grids for alignment and instrumental polarization calibration measurements.
	The IBIS spectropolarimetric $I \pm S$ images were then
	reduced applying the pipeline developed at NSO and
	slightly modified for this dataset needs,
	which takes care of dark subtraction, pixelwise gain calibration,
	co-alignment with broad-band images,
	blueshift correction, de-stretching and also the
	co-registration of all images in each scan and for the whole dataset.
	The frames were then combined and corrected for
	instrumental polarization to retrieve the Stokes vectors.
	At this stage, also the telescope induced polarization is taken into account
	\citep[as in][]{Jud10}.
	We used the closest DST telescope calibration,
	which was measured on February~16, 2007.
	
	Additionally to broad-band and narrow-band data,
	G-band filtergrams ($430.5\pm0.5$~nm) were acquired
	for a slightly smaller FOV; the pixel scale of such filtergrams 
	is $0.037''$, while the exposure time is $15$~ms.
	The fine alignment between G-band and broad-band
	images was performed through grid line targets.
	Tracking inaccuracies were corrected via a correlation procedure.
	A post processing MFBD procedure was applied also to the G-band images.
	
	As already stated, the quality of
	spectropolarimetric data has been improved by applying on
	each spectropolarimetric image the same shifts
	necessary to align and destretch the associated broad-band image
	with respect to the MFBD restored broadband image. 
	In this way,
	the seeing-induced crosstalk is reduced and a spatial resolution comparable 
	with that of the individual narrow band images is reached.
	The angular resolution of the spectropolarimetric dataset has been estimated to
	be $0.4''\simeq2$~pixels.

	The average noise level for Stokes $V$, measured as standard deviation of the
	circular polarization in continuum wavelengths,
	is $\sigma_V=3\cdot10^{-3} I_C$ (here $I_C$ is the
	continuum intensity).	
	
\section{Data analysis}
\label{Methods}
	Stokes $V$ measurements are exploited in two different ways to
	conclude on the magnetic properties of G-band bright features.
	Using the center-of-gravity method \citep[COG,][]{ReeSem79}, we calculate
	longitudinal magnetic flux density maps; the COG method is not affected
	by saturation in the kG regime, for this reason it results suitable for the analysis of
	kG bright features.
	
	Beside the COG approach, the inversion of the Stokes $I$
	and $V$ profiles of the two lines with the
	SIR code \citep{RuiCTorI92} allows us to determine the
	line-of-sight (LOS) field strength and the magnetic filling 
	factor.
	The polarization profiles emerging from the pixels under examination
	are interpreted by means of two atmospheric components: one is magnetized
	while the other is field-free. The portion of each pixel occupied by
	the magnetized component (i.e., the magnetic filling factor $f$)
	is a free parameter of the inversion. The temperature stratification
	of each component is modified with two nodes; as initial guess
	the Harvard Smithsonian Reference Atmosphere model \citep{Gin71} is used.
	The LOS velocities and the field strength in
	the magnetized component are both assumed to be constant with height. The
	stray-light contamination is considered to be unpolarized and
	is defined by averaging the Stokes $I$
	spectra in a region of $1$~arcsec around each analyzed pixel.
	Such a contribution is added to the Stokes $I$ profile synthesized from the model
	atmosphere weighted by the stray-light factor $\alpha$; the most probable value
	for the latter factor is $\alpha\simeq85\%$.
	No macroturbulence and microturbulence are considered.
	Finally, the finite spectral resolution of the instrument
	is taken into account using the known spectral PSF of the instrument
	\citep{ReaCav08}.

	Our analysis is focused on the properties of magnetic bright features.
	For this reason we analyze exclusively very strong polarization signals,
	i.e., Stokes $V$ signals above $4\times\sigma_V$ in both \ion{Fe}{1}
	lines. In spite of the severe selection rule, a total of about $53000$ pixels
	are considered; these correspond to approximately $3\%$
	of the whole FOV at any step of the time sequence and are found
	to be in strong correlation with bright features in simultaneous and cospatial
	G-band frames. On the other hand, linear polarization signals
	are found to be always below the selection threshold so that
	no full Stokes inversion is performed. In \citet{Bec07} the authors
	were able to perform full Stokes inversions in bright feature locations
	with a polarimetric sensitivity three times better than ours for the
	visible lines.
		
	To automatically identify the bright features in G-band filtergrams we adopted
	the algorithm already used by \citet{SanA07} for the identification
	of bright features in $0.2$~arcsec angular resolution G-band images filtergrams acquired at the
	DOT. The details of the procedure are explained in the paper
	cited above.
	\begin{figure*}[!ht]
  	\centering
  	\includegraphics[width=9cm]{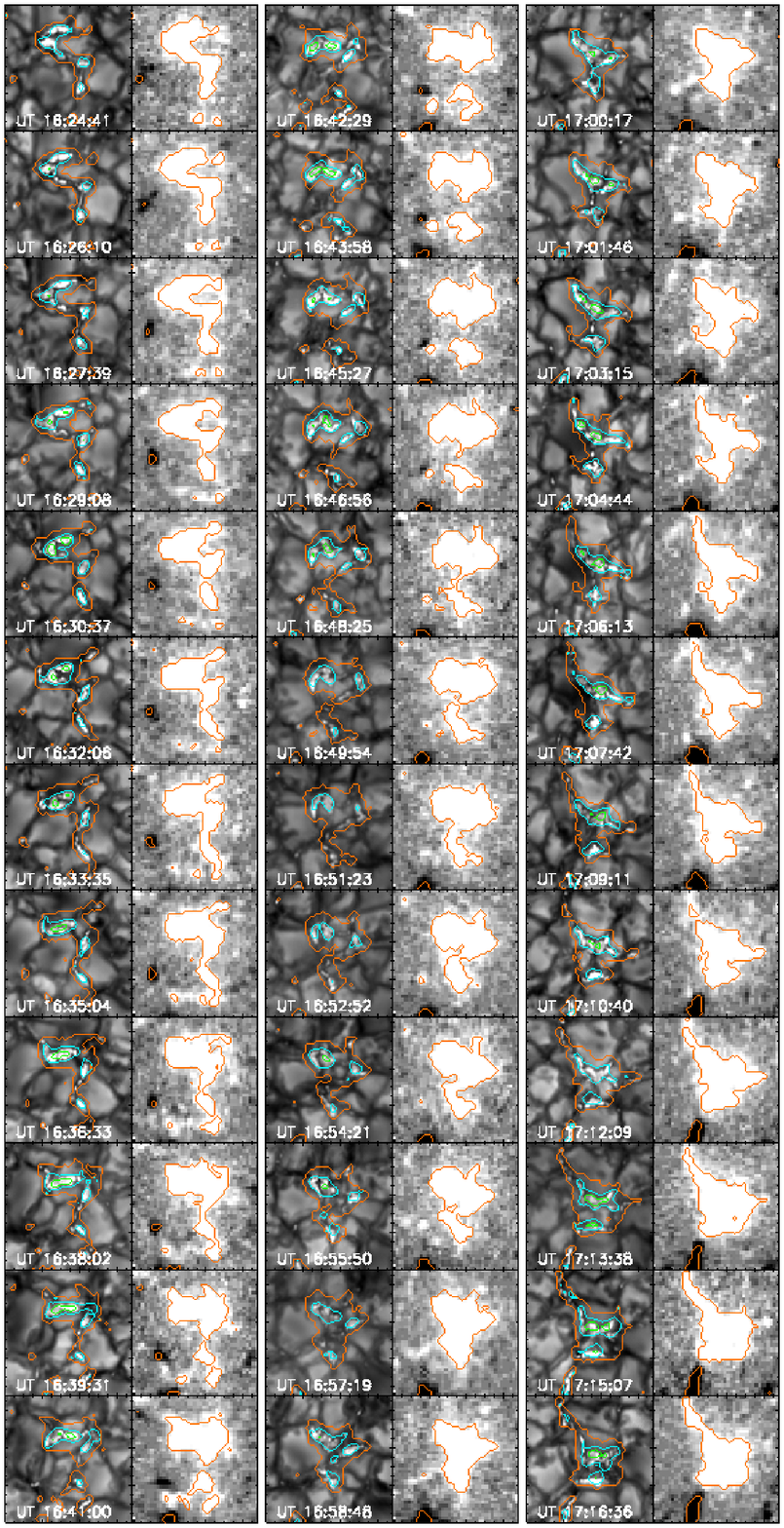}
  	\caption{Magnetic properties of a large bright feature along the whole
  		acquired time sequence. The time sequence is divided in three columns
 		representing three subsequent time intervals; in each column the time rolls from up to down.
  		For each instant of the sequence two frames are reported: 
  		a G-band image detail of the bright feature under examination (left frame) and 
		a COG magnetogram saturated to $\pm200$~G (right frame).
  		The magnetic properties as obtained from the inversion analysis are reported as
  		contours: the kG field region (orange contour), the region with magnetic filling
  		factor $>15\%$ (sky-blue contour), and the region with magnetic filling factor $>30\%$ (green contour).
  		The distance between the major ticks of the frames is of $1.4$~arcsec.\label{fig3}}
  	\end{figure*}
  	
	\begin{figure*}[!ht]
  	\centering
  	\includegraphics[width=10cm]{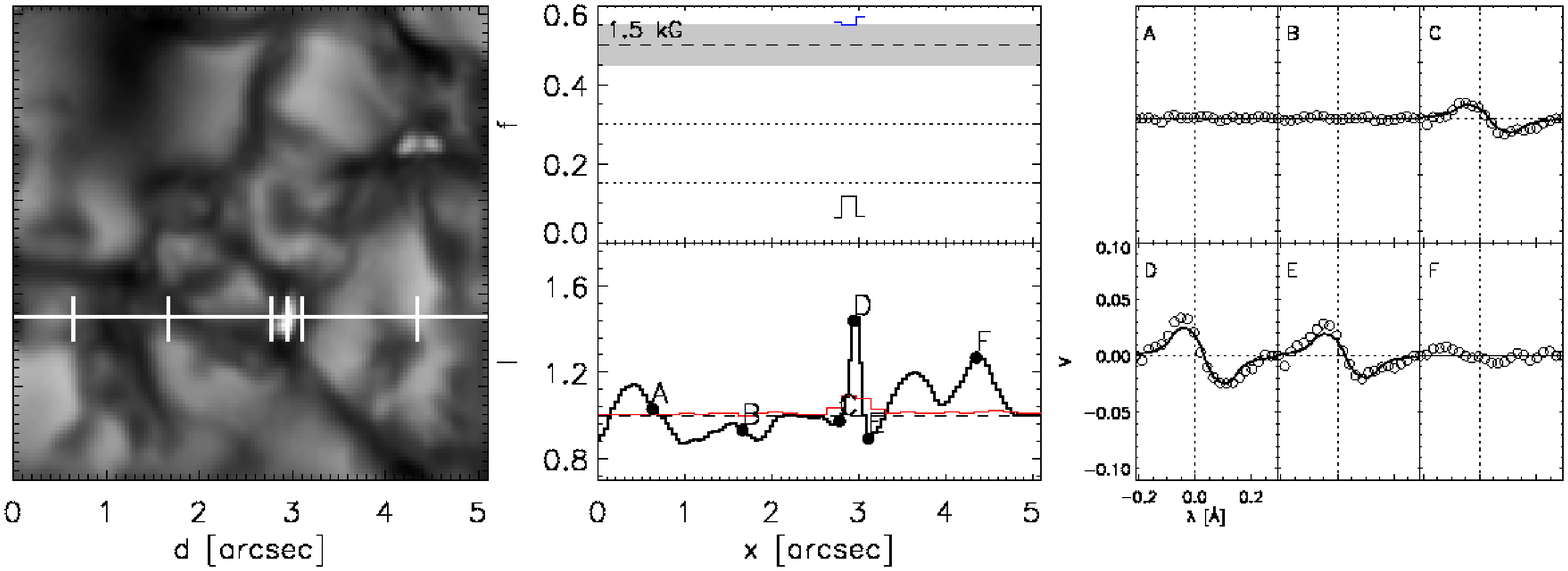}
  	\includegraphics[width=10cm]{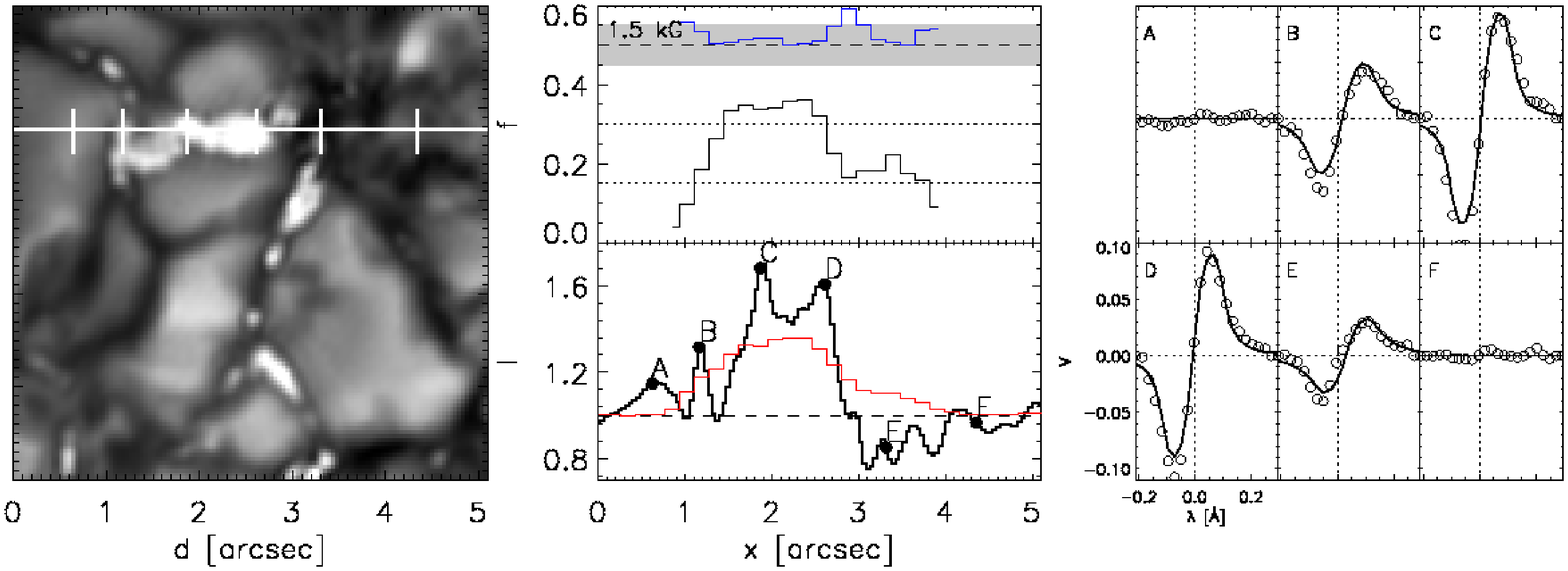}
  	\caption{Two examples of slice profiles from the results of the inversion analysis and G-band intensity
  		images. These are representative of two interesting cases, namely, 
  		a $\simeq0.3$~arcsec wide isolated bright feature observed at 16:42:29~UT (upper panel),
  		and an extended bright feature in a kG magnetized intergranular lane 
  		observed at 16:39:31~UT (lower panel, see also Fig.\ref{fig3}).
  		Left panel: a $d\times d$ G-band intensity detail image to identify the position of
  		the slice here represented as a white line (left image).
  		Central panel, upper plot: magnetic filling factor $f$ derived from the inversion analysis (solid black line),
  		the horizontal dotted lines represent the levels defining the filling factor contours
  		in Fig.~\ref{fig3} (i.e., $f=15$\% and $f=30$\% levels, respectively); here we plot
  		also the magnetic field strength derived from the inversion (blue line) whose variation refers to
  		the $1.5$~kG level represented as a dashed line; the shaded area
		represent a $\pm150$~G field strength variation around the $1.5$~kG level.
		Central panel, lower plot: G-band intensity along the slice
  		normalized to the average intensity of the time
  		sequence (black line) and $\vert V \vert$ maximum amplitude along the slice (red line) 
  		properly rescaled to be compared with the G-band intensity profile. The histogram-like
  		representation adopted in these plots allows us to highlight the different pixel
  		scales of G-band data ($0.037$~arcsec) and spectropolarimetric data ($0.18$~arcsec), respectively.
  		Right panel: \ion{Fe}{1} $6301.5$~\AA{} Stokes $V$ profiles from both observation and
  		inversion analysis (right panels, symbols and solid line, respectively) picked at different positions
  		along the slice (named as $A$, $B$, $C$, $D$, $E$, and $F$); each profile is
  		represented in the $x-y$ range specified in the plot $D$ (wavelengths are represented as
  		deviation with respect to the laboratory wavelength of the line). Positions $A$, $B$, $C$, $D$, $E$, and $F$
  		are reported in the G-band detail image as ticks and on the intensity profile with dots.\label{fig4}}
  	\end{figure*}
  	
  	\begin{figure}
  	\centering
  	\includegraphics[width=8cm]{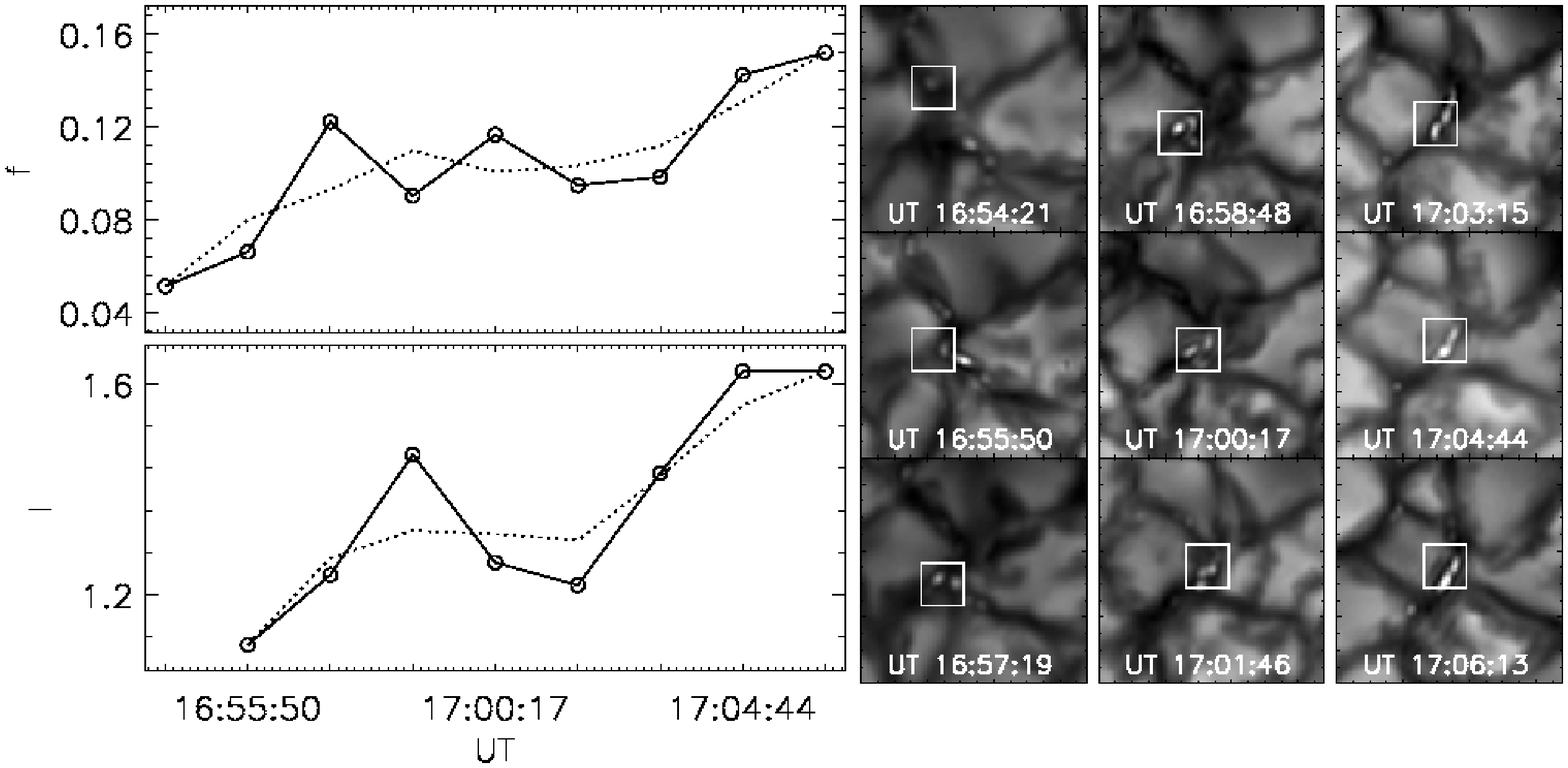}
  	\caption{Temporal variation of both brightness and magnetic filling
  		factor in a process of merging of bright features. Left panel, upper plot: temporal variation
	  	of the magnetic filling factor. Left panel, lower plot: temporal variation
	  	of bright feature brightness (normalized to the average intensity of the
	  	time sequence). In both the plots the dotted line represents the
	  	trend of the evolution obtained by smoothing the temporal
	  	variation represented with the solid line. 
	  	In the right panel the time sequence is divided in three columns
 		representing three subsequent time intervals; in each column
 		the time rolls from up to down. Here, G-band image details of the bright feature
	  	under examination are reported. The square indicates the region from which
	  	the quantities in the left panel are derived; this is placed where the
	  	strongest Stokes $V$ signal is found in
	  	the region under examination. At UT~16:54:21 no bright
		features were found by the automatic procedure in the
		selected square.\label{fig8}}
  	\end{figure}
	  
\section{Results}
\label{Results}
	We investigated the magnetic properties of G-band bright features
	through both COG magnetograms and inversion of Stokes
	$V$ profiles. In the following we present the results derived from
	such analyses in two separated sections.
	The first one (\S~\ref{ResultsGbMag}), concerning the analysis
	of magnetograms, allows us
	a straightforward comparison with the results presented in the
	extended literature on this topic (see \S~\ref{Intro}). The second one
	(\S~\ref{ResultsGbInv}),
	concerning the inversion analysis, offers new insight on the
	relation between photometric and magnetic properties of G-band
	bright features sheding new light on the results
	in \S~\ref{ResultsGbMag}.
	
\subsection{G-band intensity and COG maps}
\label{ResultsGbMag}
	Fig.~\ref{fig1} (upper panel) shows an
	example of bright features
	identified through the automatic procedure described in
	\S~\ref{Methods}
	on one of the snapshots of the sequence.
	The average fraction of solar photosphere occupied by such features
	during the time sequence is about $0.6$\%.
	Their linear dimensions range between $\simeq0.1$~arcsec  up to $2-3$~arcsec. 
	They appear in a great variety of shapes: from the point-like
	(at the angular resolution of the G-band data) to the extended
	and elongated (e.g., the feature in the center of the FOV
	in the upper panel of Fig.~\ref{fig1}).
	
	In Fig.~\ref{fig2} (left panel) we report the scatter plot relating
	the average G-band intensity of bright features $\bar{I}$ (normalized to the
	average intensity of the time sequence) and their area $A$.
	The apparent linear relation between $\bar{I}$ and $A$ is
	broken for $A>0.1$~arcsec$^2$ and $\bar{I}$ seems to saturate
	for larger $A$ values.
	Hence, we fitted an exponential function $a_1\cdot e^{b_1x}+c_1$
	on the scatter plot, where $a_1$, $b_1$, and $c_1$
	are free parameters of the fit. The exponential fit is overplotted on
	the data while the derived parameters are reported in Table~\ref{tabfit}.
	According to the exponential fit, we can define a
	saturation level for the average intensity of large bright features
	approximatively at $c_1=\bar{I}_{sat}=1.4$.
	In the plot $\bar{I}_{sat}$ is represented as a horizontal
	dashed line, while the shaded area represents the error
	on $\bar{I}_{sat}$ derived from the fit procedure.	
	
	In Fig.~\ref{fig2} (central panel) we report a scatter plot relating
	the average COG magnetic flux density $\bar{B}_{mag}$
	and $\bar{I}$ of identified bright features.
    In order to compare these quantities,
	we rescaled the COG magnetograms
	(Fig.~\ref{fig1} lower panel)
	at the G-band pixel scale, so that we could
	associate to each identified feature
	its $\bar{B}_{mag}$ and $\bar{I}$
	computed over the pixels forming the region
	itself. We find that also the relation
	between $\bar{B}_{mag}$ and $\bar{I}$ can be described by
	an exponential law.
	The fit displayed in Fig.~\ref{fig2} (central panel)
	was obtained by imposing the saturation level to be $\bar{I}_{sat}$
	(i.e., $a_2\cdot e^{b_2x}+\bar{I}_{sat}$).
	The values for $a_2$ and $b_2$ derived from the fit
	are reported in Table~\ref{tabfit}.

	In the rightmost panels of Fig.~\ref{fig2} we
	report the histograms for $\bar{B}_{mag}$
	and $\bar{I}$.
	The most probable value for the average flux density in bright features is
	approximatively $100-150$~G with a tail up to $\simeq800$~G, while the normalized average
	intensity ranges between $0.9$ and $1.4$ with the most probable value
	at $\simeq1.1$. 
	
	Both the scatter plots in Fig.~\ref{fig2} have been defined
	by considering the data from the whole time sequence.
	It is important to specify that both plots are good representations
	of the relations between the magnetic flux density, the G-band intensity,
	and the area of bright features at any instant of the time sequence.
	This has been checked by comparing the behavior of all $36$ scatter plots
	taken individually along the time sequence with the scatter plots
	of Fig.~\ref{fig2}. In Fig.~\ref{fig7} we show
	one of the results of such a check; we report the $\bar{I}$ vs. $A$
	scatter plot for each instant of the time sequence in comparison with
	the results of the exponential fit in Fig.~\ref{fig2} (left panel).
	We have to specify that in Fig.~\ref{fig7} $\bar{I}$
	is obtained by normalizing the intensity to the average G-band
	intensity of each frame. The agreement between the scatter plots and
	the exponential behavior derived from the time-integrated scatter plot in
	Fig.~\ref{fig2} is more than satisfactory.
	Once the consistency has been verified,
	we are allowed to use the time integrated scatter plot, as it provides
	a more statistically sound representation.
	\begin{figure}[!ht]
  	\centering
  	\includegraphics[height=7cm]{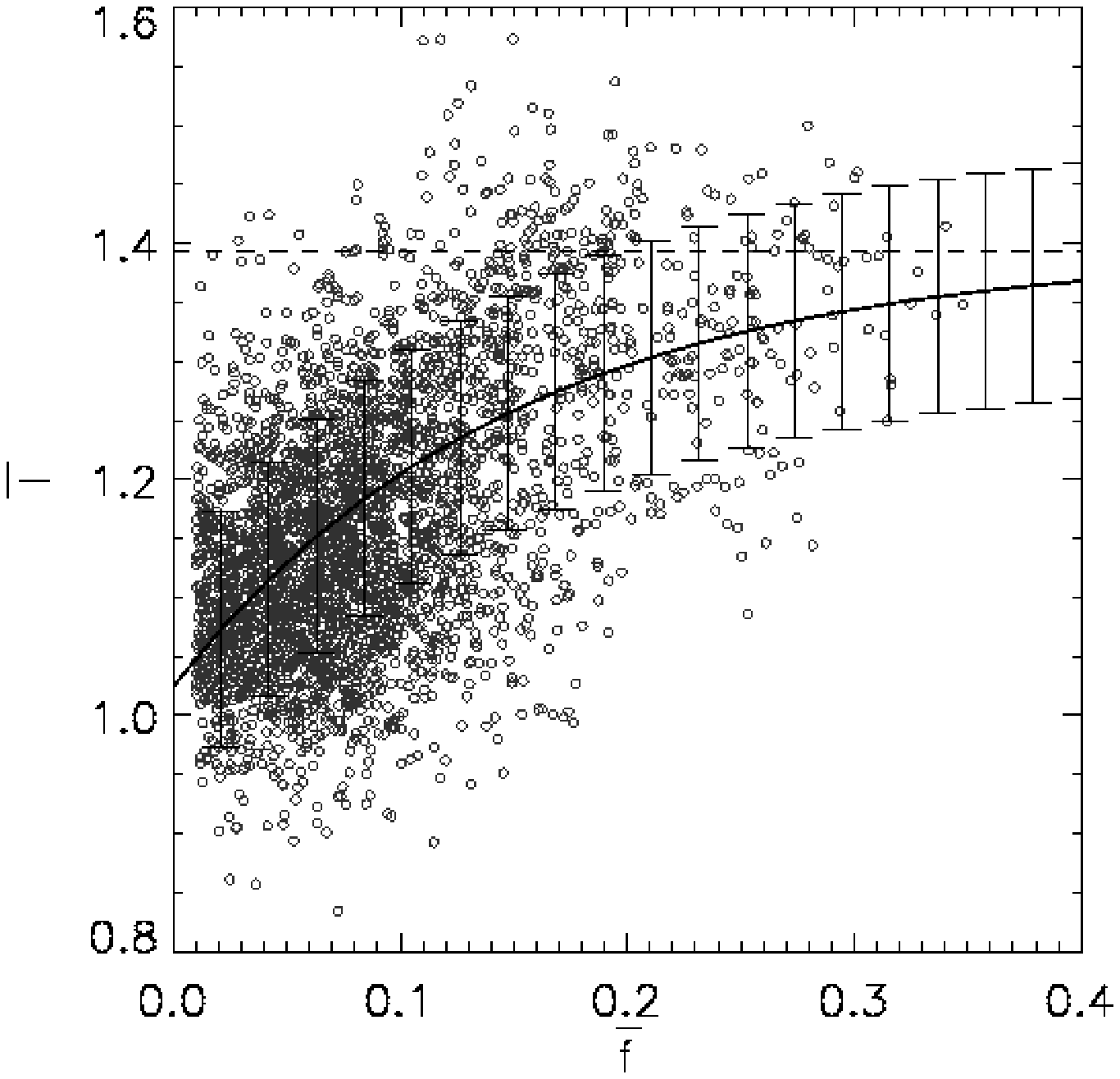}
  	\caption{Correlation between average magnetic filling factor $\bar{f}$ and 
  		average G-band intensity $\bar{I}$ (normalized to the average intensity of the time sequence)
  		for bright features in the analyzed dataset. In the
		plot an exponential fit of the data is reported (solid line).
		The error bars represent the average standard deviation error of the fit;
		the horizontal dashed line represent the saturation value $\bar{I}_{sat}$.
		For details on the fit procedure refer to \S~\ref{ResultsGbMag}, \S~\ref{ResultsGbInv}, and
		Table~\ref{tabfit}.\label{fig5}}
  	\end{figure}

\subsection{G-band intensity and inversion results}
\label{ResultsGbInv} 	
  	Fig.~\ref{fig3} shows the temporal evolution of the magnetic properties of the largest
  	bright feature in the FOV, as derived from the inversion analysis, together with
  	G-band and COG magnetogram images. The feature can be
  	considered a small network patch, i.e., a region in which the plasma dynamics
  	concentrates kG fields over a region $2-3$~arcsec wide.
  	kG fields are revealed over the entire
  	region comprising the bright feature (orange contour).
  	High magnetic filling factors are found
	in correspondence with the brightest portions
	of the feature. Namely, in these regions $f$ is mostly higher than
  	$15$\% and can reach values up to $\simeq35$\%.
	Such values are much higher than typical 
  	$f$ values for kG fields in the internetwork which
  	have been measured to be of few percents \citep[e.g.,][]{Vit10}.
  	This correlation is persistent along the whole time 
  	sequence (cfr., the different snapshots
	of Fig.~\ref{fig3}).
   	
  	Fig.~\ref{fig4} shows the pixel-by-pixel
  	correlation between Stokes $V$ amplitude,
  	magnetic filling factor $f$, and G-band intensity $I$ (normalized to the
  	average intensity of the whole time sequence)
  	along slices extracted from the analyzed dataset.
  	This representation is similar to the one adopted
  	by \citet{Ber04} to represent results from
  	observations performed in different bands.
	Stokes $V$ profiles from both observations and the inversion analysis 
	for selected positions along the slices are also represented in the figure.
	These profiles are well fitted by the SIR code
	under the adopted inversion hypotheses (\S~\ref{Methods}).
  	In the two panels of the figure the variation
  	of the G-band intensity is found to be in correlation
  	with both Stokes $V$ amplitude and $f$, while the
  	magnetic field strength does not show any correlation with
  	these quantities; in fact, it slightly varies around $1.5$~kG.
	The two cases reported in Fig.~\ref{fig4} are representative
	of a point-like bright feature in an intergranular lane (upper panel),  
	and an extended bright feature in an intergranular lane (lower panel).
	
	The upper panel of Fig.~\ref{fig4} displays the different capabilities
	of G-band and spectropolarimetric data
	in resolving the properties of bright magnetic features.
	In fact, the bright feature is found to be $\simeq0.15$~arcsec wide,
	i.e., four pixels in G-band data, while
	the polarization signal is spread over a region
	of $\simeq0.7$~arcsec, i.e. four pixel in spectropolarimetric data.
	This is a good example of a magnetic feature whose dimension is close to
	the limit of the angular resolution of spectropolarimetric data (\S~\ref{Data}).
	According to the criterion described in \S~\ref{Methods},
	we inverted the profiles of three of these four pixels,
	i.e., positions $C$, $D$, and $E$
    where we measured approximatively $1.5$~kG,
	and a filling factor approximately of $10$\%.
	
	In the lower panel of Fig.~\ref{fig4}
	a horizontal slice is extracted in correspondence
	with a strongly magnetized intergranular lane.
	The increase of brightness
	of the feature from position $B$ to $D$
	is correlated with the increase of the magnetic filling
	factor, that reaches values above $30$\%.
	The Stokes $V$ amplitude variation measured in positions
	$B$, $C$ and $D$ is also evidently correlated with
	the G-band intensity variation found along the slice.
	The plots also show that quite high values for both magnetic filling factor and magnetic 
	field intensity are found in the dark part of the intergranular lane (positions $E$ to $F$).
	In particular, we notice that at positions $B$ and $E$ the estimated filling factor values are
	similar (about 20$\%$), while the 	
	observed G-band intensities are quite different. 
	This difference has to be ascribed to the spurious
	contributions to the polarimetric signal in pixel $E$ from the adjacent 
	magnetized pixels (the two bright features right at the left and below position $\simeq[3,3]$~arcsec
	in the G-band subfield).
		
	The correlation between the magnetic filling factor
	and the G-band intensity shown in Figs.~\ref{fig3}~and~\ref{fig4}
	is found for all the bright features
	observed along the time sequence and selected by the automatic procedure.
	As a further example, in Fig.~\ref{fig8} we report the temporal variation
	of both brightness and magnetic filling factor in
	a process of merging of faint bright features in a
	single high brightness feature. To derive the plots in
	such a figure we focused on the properties of the pixels
	around the strongest Stokes $V$ signal in the selected subfield
	(white square on G-band images).
	This choice was necessary since for the first five instants
	many distinct bright features were recognized by the automatic
	procedure and this does not allow one to define a unique temporal
	evolution. From the plot we can recognize a correlation
	between the increase of brightness and magnetic filling factor.	 
	
	In Fig.~\ref{fig5} such a correlation is made explicit by
	the scatter plot relating
	$\bar{I}$ and the average filling factor
	$\bar{f}$ calculated over the features selected by
	the automatic procedure. Also in this scatter plot we
	report the result of an exponential fit performed
	adopting the function $a_3\cdot e^{b_3x}+\bar{I}_{sat}$.
	The values of $a_3$ and $b_3$ derived from the fit
	are reported in Table~\ref{tabfit}.
	
	A simple test of the consistency of the results obtained
	from the fits can be done by comparing the values of
	$b_2$ and $b_3$ (second and third lines of Table~\ref{tabfit}).
	In fact, if we consider $1.7$~kG as the typical field strength
	value derived from the inversions, we find that the values
	reported in Table~\ref{tabfit} satisfy $b_2\simeq b_3/(1.7\times10^3$~G).
	This relation stems from the fact that $b_2$ and $b_3$ are the
	inverse of a magnetic flux density and a magnetic filling factor,
	respectively.
		
\section{Discussion}
\label{Disc}
	The automatic identification algorithm applied on G-band filtergrams
	allows us to estimate that $\simeq 0.6$\% of the dataset FOV is occupied by bright features.
	This result is in good agreement with the value reported by \cite{SanA04}
	for the quiet photosphere observed at $0.14$~arcsec angular resolution.
	
	As expected, the SIR inversion retrieves kG fields where
	G-band bright features are found \citep[e.g., ][]{She04}.
	This co-spatiality, here presented only for three examples of features in the FOV,
	is common to all the bright features individuated in the dataset
	\citep[see][for three more examples of small features]{Vit09}.
	
	The spectropolarimetric data have a PSF with a
	FWHM of $\simeq0.4$~arcsec, as can be deduced by
	the shape of the $\vert V \vert$ maximum amplitude showed in
	the upper panel of Fig.~\ref{fig4}.
	Therefore, the low filling factor retrieved by the
	spectropolarimetric inversion can be ascribed to the
	intrinsic size of the magnetic counterparts of the
	G-band bright features and/or to the PSF shape.
	Even considering the several effects which can
	contribute to a PSF with extended tails, namely,
	the part of the incoming aberrations not corrected by the AO,
	scattered light in the optical path or by the atmosphere,
	and defects in the image calibration pipeline,
	it seems unreasonable that the PSF effect is the
	dominant one, as we should have a PSF with extremely extended tails.
	Moreover, for the case represented in the upper
	panel of Fig.~\ref{fig4}, the filling factors found are compatible
	with a magnetic feature almost filling a single IBIS pixel
	($0.18$~arcsec) spread by a $0.4$~arcsec
	FWHM wide PSF.	
	Therefore, we interpret the $\simeq15-30$\%
	filling factors found for the extended bright feature
	(Fig.~\ref{fig3}~and lower part of Fig.~\ref{fig4})
	as a signature of unresolved magnetic features.
	A deeper analysis of the PSF convolution effects on
	spectropolarimetric inversions will be the topic
	of a forthcoming paper.

	From the unsigned flux density histogram, we can infer that the
	most probable unsigned flux density of the bright features is
	approximatively $100-150$~G (Fig.~\ref{fig2}, right panels). 
	This value is about twice the value obtained by \citet{Bec07},
	but	is in agreement with \citet{BerTit01}.
	The disagreement among the results can stem from the different
	spatial resolution of the magnetograms: the spectropolarimetric data
	employed in \citet{Bec07} had a spatial resolution of about $1$~arcsec,
	while the magnetograms used in \citet{BerTit01} had a $0.3$~arcsec
	spatial resolution, much closer to that of our dataset.

	When plotting the G-band intensity vs. the magnetic flux density
	(Fig.~\ref{fig2}, central panel), we find that the G-band intensity
	seems to saturate for values of $\bar{B}_{mag}>500$~G.
	We can compare this plot with Fig.~12 of \citet{BerTit01} and the
	G-band contrast of identified magnetic brightening as a function
	of the peak magnetic flux density in Fig.~9 of \citet{Ber07}, with the
	caveat that the authors considered peak quantities instead of average quantities.
	The scatter plot relating the G-band contrast and the total
	integrated polarization reported by \cite{Bec07} also shows an
	evident saturation of the contrast around $0.53$ for the
	reported values of the total polarization.
	In those plots, even if the contrast values are in agreement
	with the ones we measured, the authors did not find any relation between the two quantities.
	We can note, however, that our statistics is
	much richer than in \citet{Bec07}. Such a difference
	in the statistics stems from two main facts. The first is
	that we analyzed $35$ sets of co-spatial and
	co-temporal G-band and spectropolarimetric data, while
	in \cite{Bec07} $8$ sets were available. Moreover, different
	procedures for the selection of bright features were
	adopted; in \citet{Bec07} the authors specify that their
	procedure usually leads to the selection of extended
	patches that could be divided in smaller ones, this
	caused an important reduction of the total number of
	selected bright features.
	
	In particular, the latter two works lack statistics exactly for
	low magnetic flux densities and try to fit their scatter plots with
	linear functions, while the former, on the contrary,
	only shows magnetic flux densities below $400$~G.

	Commenting the plot in Fig.~\ref{fig2} (left panel), the increase of G-band
	intensity with the increase of the size of bright features is in general agreement
	with previous results from high spatial resolution observations
	\citep[e.g.,][]{Ber95,BovWie03,Wie04,Hirz05,Ber07}.
	The G-band contrast saturation values reported are usually between
	$1.2$ and $1.4$ for disk-center observations, nevertheless, due
	to the large scatter of data, different interpretations of the results have been suggested.
	In \citet{Wie04} and \citet{Ber95} for instance, the increase
	was within the dispersion of the results and the authors concluded
	that no clear trend of intensity as a function of size was observed,
	the contrast being essentially constant.
	\citet{Ber07} showed instead that their results can be fitted by a straight line. 
	\citet{BovWie03} reported a clear increase of contrast with
	size and fitted the results with a polynomial function of third degree.
	The results presented in \citet{Utz09} are an exception in this context,
	since the authors reported a decrease of contrast for
	bright features larger than \textit{HINODE} resolution.
	The authors interpreted this trend as an effect of the
	segmentation algorithm employed to detect the analyzed features.
	Close inspection of all these data \citep[with the exception of the ones in ][]{Utz09}
	reveals an increase of contrast when the size of the feature is
	comparable with the dataset resolution and a saturation at the largest sizes.
	This is in agreement with our finding and we can notice that even
	the threshold spatial scale value of contrast saturation that we
	obtain from our analyses is compatible with previously presented results.
	In fact, assuming radial symmetry for the selected features,
	we find a threshold size value of approximately
	$2\times\sqrt{1/(\pi b_1)}\simeq0.3$~arcsec~$\simeq210$~km,
	in agreement with \citet{Ber95}, \citet{Wie04}, and \citet{Ber07},
	when taking into account the different spatial resolution of the data.
	
	This contrast/size relation is in apparent disagreement with classical
	flux tube models, according to which contrast decreases with
	the size of the magnetic feature \citep[e.g.,][]{Spr1976,Dein1984,Pizzo1993}.
	Recently, \citet{CrisRas09} showed that this effect can be
	interpreted as a signature of unresolved magnetic features. 
	The authors, using a two-dimension numerical model of isolated
	and clustered magnetic flux tubes, simulated the emergent
	intensity for different cluster dimensions at different angular resolutions.
	They found an exponential like relationship between size and
	continuum contrast, which saturated for cluster diameters larger
	than twice the spatial resolution of the observations.
	This picture is also corroborated by the fact that the finite spatial
	resolution of observations mostly affects small size features
	\citep[e.g.,] [and references therein]{CrisErm08}: the smaller the
	feature the larger is the reduction of its contrast with the decrease of spatial resolution.
	Larger features are instead less affected, thus explaining
	the general agreement of contrast values presented in the literature at the saturation.
	
	The work of \citet{Wie04}, which showed that both the $588$~nm
	continuum contrast and the G-band contrast of bright features
	saturate at the same spatial scale, allows us to directly compare
	our result with \cite{CrisRas09} and the two results seem to match quite congruously.
	In fact, we found the saturation for $A>1/b_1\simeq0.07$~arcsec$^2$ (Table~\ref{tabfit}),
	for which we can calculate an equivalent diameter of $d\simeq0.3$~arcsec,
	i.e. about twice the spatial resolution of our G-band dataset.

	The plot reported in Fig.~\ref{fig5}, showing the relation between magnetic
	filling factor and G-band intensity found in our data, also supports the
	idea of the presence of unresolved magnetic features in our observations.
	We remember that, only the Stokes profiles cospatial with pixels which show
	G-band brightness enhancement have been considered to define the plot and that the angular
	resolution of the spectropolarimetric data has been estimated to be $\simeq0.4$~arcsec.
	From these inversions, it follows that filling factor values are mostly
	$\lesssim10$\% and only a small fraction of the bright features present filling factors above $20$\%.
	Beside this, the G-band filtergrams have a spatial resolution of $\simeq0.1$~arcsec,
	therefore in these images we can resolve bright features that are as small
	as $25$\% of the spectropolarimetric data pixel.
	The low values of the filling factor suggest that the smallest
	scale bright features are smaller than $0.1$~arcsec.
	
	The variation of the G-band intensity for point-like features
	($A\lesssim0.1$~arcsec$^2$) supports this idea: fully resolved bright
	features should not change their intensity with their dimension.
	We recall that, in G-band bright features, the brightness
	enhancement occurs because of the local evacuation induced by the high magnetic pressure.
	We can assume that increasing the magnetic filling factor means
	enlarging the evacuated region in our pixel and therefore interpret
	the plots of Figs. \ref{fig3}, \ref{fig8}, \ref{fig4} and \ref{fig5} with the consequent
	enhancement of the number of photons that emerge from hot deep layers
	in the photosphere in the pixels forming the bright feature under examination.
	These arguments, combined with the G-band intensity saturation found for
	$A\gtrsim0.1$~arcsec$^2$ and $f\gtrsim20$\%, suggest that the larger G-band
	bright features can be thought as clusters of kG ``elementary
	bright points'' having typical dimension $\lesssim0.1$~arcsec.
	This is in full agreement with \citet{CrisRas09}.
	
	From our inversion analysis we conclude on the role of the magnetic
	filling factor of kG fields in the formation of G-band bright features
	(Fig.~\ref{fig5}): the larger the magnetic filling factor the brighter the feature.
	To explain our findings, we can refer to \cite{deW05}.
	The authors commented about the observation of G-band bright
	features in quiet Sun (inter-network) regions: \textit{``we conclude that,
	even though the inter-network bright points may momentarily become invisible,
	the magnetic field element remains and may become bright again at some later time.
	This agrees well with the conclusion of \cite{BerTit01} that magnetism is
	a necessary but not a sufficient condition for the formation of a network bright point''}.
	The results here presented in \S~\ref{Results} complement their
	argumentation. In terms of the spatial extension of kG flux
	tube clusters the sentence can be reworded as:
	the larger the cluster the brighter it will appear, but, once the
	cluster dimensions are comparable with the spatial resolution
	of the observations, the brightness saturates.
	From this follows that small G-band bright features can be
	momentarily invisible simply because the local spatial organization 
	of kG fields is temporarily changed, for example, through the action of the
	photospheric dynamics or because the spatial resolution decreases due to seeing.
	This affects the strongest the brightness of point-like features
	\citep[as already shown in ][]{Vit09}, but also can induce the internal
	brightness variation of large features (Figs.~\ref{fig3} and \ref{fig4}).
	In the recent \citet{SanA10} the measure of bright feature abundance
	in the quiet Sun is based on the concept that small scale faint
	features can be extremely variable in time.
	Applying an identification procedure that makes use of $0.1$~arcsec
	angular resolution G-band filtergrams time sequences, the authors
	were able to rise the estimate of G-band bright
	feature abundance of a factor three, i.e., up to $\simeq1$ BPs every $2$~arcsec$^{2}$.
	
	It is worth also to focus
	on recent alternative interpretations of the magnetic structure of
	photospheric bright features. In the last decade many works
	based on modern MHD simulations performed with the MURaM code
	have been dedicated to the study of G-band bright features
	\citep[][]{Sch03,She04,Vog05,She07}. These were able to
	reproduce the photometric properties of G-band observations
	and to confirm the association between local brightness enhancement
	and small scale kG concentrations. In all these works a common
	figure for the magnetic structure of bright features emerges:
	at $20$~km spatial resolution these are ``elongated sheet-like''
	features. No evidence of internal structuring, e.g., through
	many point-like features is found. In agreement with this
	interpretation, \citet{Ber04} reported that no evidence of
	internal structuring was found for bright
	features analyzed in both $0.1''$ resolution G-band images
	and $0.18''$ resolution magnetograms
	obtained at SST. In contrast with our interpretation,
	the authors put forward that an internal variation
	of brightness of extended bright features can be
	explained via an internal variation of field strength
	instead of magnetic filling factor.
	In the recent work of \citet{Nar10} the authors
	analyzed full Stokes measurements performed through
	CRISP at Solar Swedish Tower (SST) with $0.15$~arcsec angular resolution.
	On the same line of \citet{Ber04}, the authors analyzed
	Stokes $V$ profiles adopting a Milne-Eddington code and
	by imposing the magnetic filling factor to be equal to
	unity, i.e., considering magnetic flux
	concentrations to be resolved in their observations.
	In their work the authors properly discussed this choice
	specifying that their measure of the magnetic field should
	be considered either as a measure of the magnetic flux density
	\citep[as in][]{Ber04} or as a measure of a roughly defined
	average of the magnetic field strength.
	On the other hand, our approach considers the effect of the limited spatial
	resolution in the inversion analysis of $0.4$~arcsec
	spatial resolution data; this allows us
	to separate the contribution of
	the field strength and the magnetic filling factor.
	
	The works of \citet{Nar10}, based on CRISP data, 
	\citet{Dan10b} and other observations performed with SUNRISE/IMaX, 
	represent the first effort in trying to understand the
	photospheric magnetism at $0.15''$ angular resolution.
	Such observations will shed new light in the study of the
	structuring of the photospheric magnetic field.
	
	As a conclusive part of our discussion we
	want to point out two recent works which
	support a fine structuring of photospheric bright
	features. In the first one, \citet{Goo10}, the observation
	of point-like bright features in $0.12''$ resolution
	observations at $705.7$~nm is reported, while no sheet-like features were
	observed. The second one is the work of \cite{Dan10} in which
	processes of magnetic intensification
	are simulated via the MURaM code. Such processes form
	stable point-like (i.e., $\simeq0.1''$ wide) kG
	features which are able to produce
	a continuum brightness enhancement at $630$~nm.
	These are different from the strong sheet-like kG
	concentrations produced by advecting field lines
	in downflow regions \citep[e.g.,][]{Sch03}. From our
	point of view, it could be extremely interesting
	an in-depth study of the spatial and temporal evolution
	of such kG elements to check whether they can produce,
	by grouping, extended $\sim1''$ wide bright
	features.
	
\section{Conclusions}
\label{Conc}
	We have presented new results from the analysis
	of observations with IBIS in spectropolarimetric mode.
	These are, at the same time, complementary and
	corroborative of the results reported in \citet{Vit09}.
	Namely, in the present work we confirm the correlation
	between bright feature brightness and kG magnetic
	filling factor found in the analysis of three particular
	cases of temporal evolution of small-scale features in
	\citet{Vit09}: the higher the local kG filling
	factor the brighter the G-band feature.
	This conclusion, first evinced by the detailed
	analysis of an extended bright feature (i.e., a small network patch),
	has been validated by a statistical study of the properties
	of all the bright features observed in the FOV at any
	instant of the time sequence.

	Considering the angular resolution of our spectropolarimetric
	($\simeq0.4$~arcsec) and G-band ($\simeq0.1$~arcsec) data, we put forward
	the upper limit of $\simeq0.1$~arcsec for the typical dimension over which the
	``elementary G-band bright features'' (i.e., bright points) are formed.
	Adopting this figure, larger G-band bright features could be
	thought to be clusters of bright points
	\citep[in agreement with][]{CrisRas09}. Such substructuring
	of bright features can be just inferred from $0.4$~arcsec
	angular resolution IBIS spectropolarimetric data. New
	higher angular resolution observations will offer to
	the solar community the opportunity to conclude on
	the magnetic structure of G-band bright features.
	
	\acknowledgements
	We are very grateful to the anonymous referee for the
	constructive comments and remarks on the manuscript.
	The authors acknowledge the contribution of Luis Bellot Rubio
	and Alexandra Tritschler in the inversion analysis and
	in the reduction/reconstruction of data.
	This work was partially supported by the MAE Spettro-Polarimetria Solare 
	Bidimensionale research project, by the
	Agenzia Spaziale Italiana through grant ASI-ESS, by the
	Istituto Nazionale di Astrofisica through grant PRIN-INAF 2007.
	NSO is operated by the Association of Universities for Research 
	in Astronomy, Inc. (AURA), under cooperative agreement with the 
	National Science Foundation.
	The authors are grateful to the DST observers D. Gilliam, M. Bradford and
	J. Elrod. IBIS was built by INAF-Osservatorio Astrofisico 
	di Arcetri with contributions from the Universit\`a
	di Firenze and the Universit\`a di Roma ``Tor Vergata''. 
	The authors acknowledge F. Cavallini, K. Reardon, 
	and the IBIS team for their invaluable and unselfish support.

%---- Bibliography
	%\bibliography{Bib_IBIS_5}

\begin{thebibliography}
	%\expandafter\ifx\csname natexlab\endcsname\relax\def\natexlab#1{#1}\fi

	\bibitem[{{Beck} {et~al.}(2007){Beck}, {Bellot Rubio}, {Schlichenmaier}, \&
  		{S{\"u}tterlin}}]{Bec07}
		{Beck}, C., {Bellot Rubio}, L.~R., {Schlichenmaier}, R., \& {S{\"u}tterlin}, P.
  		2007, \aap, 472, 607

	\bibitem[{{Berger} {et~al.}(1995){Berger}, {Schrijver}, {Shine},
  		{et~al.}}]{Ber95}
		{Berger}, T.~E., {Schrijver}, C.~J., {Shine}, {et~al.} 1995, \apj, 454, 531

	\bibitem[{{Berger} \& {Title}(1996)}]{BerTit96}
		{Berger}, T.~E., \& {Title}, A.~M. 1996, \apj, 463, 365

	\bibitem[{{Berger} \& {Title}(2001)}]{BerTit01}
		---. 2001, \apj, 553, 449

	\bibitem[{{Berger} {et~al.}(2004){Berger}, {Rouppe van der Voort},
  		{L{\"o}fdahl}, {Carlsson}, {Fossum}, {Hansteen}, {Marthinussen}, {Title}, \&
  		{Scharmer}}]{Ber04}
		{Berger}, T.~E., {Rouppe van der Voort}, L., {L{\"o}fdahl}, M.~G.,
  		{Carlsson}, M., {Fossum}, A., {Hansteen}, V.~H., {Marthinussen}, E., {Title},
  		A., \& {Scharmer}, G. 2004, \aap, 428, 613

	\bibitem[{{Berger} {et~al.}(2007){Berger}, {Rouppe van der Voort}, \&
  		{L{\"o}fdahl}}]{Ber07}
		{Berger}, T.~E., {Rouppe van der Voort}, L., \& {L{\"o}fdahl}, M. 2007, \apj,
  		661, 1272
  
	\bibitem[{{Bharti} {et~al.}(2006){Bharti}, {Jain}, {Joshi}, \&
  		{Jaaffrey}}]{Bha06}
		{Bharti}, L., {Jain}, R., {Joshi}, C., \& {Jaaffrey}, S.~N.~A. 2006, in
  		Astronomical Society of the Pacific Conference Series, Vol. 358, Astronomical
  		Society of the Pacific Conference Series, ed. {R.~Casini \& B.~W.~Lites},
  		61--+

	\bibitem[{{Bovelet} \& {Wiehr}(2003)}]{BovWie03}
		{Bovelet}, B., \& {Wiehr}, E. 2003, \aap, 412, 249

	\bibitem[{{Bovelet} \& {Wiehr}(2008)}]{BovWie08}
		---. 2008, \aap, 488, 1101

	\bibitem[{{Caccin} \& {Severino}(1979)}]{Cac79}
		{Caccin}, B., \& {Severino}, G. 1979, \apj, 232, 297

	\bibitem[{{Cavallini}(2006)}]{Cav06}
		{Cavallini}, F. 2006, \solphys, 236, 415

	\bibitem[{{Criscuoli} \& {Ermolli}(2008)}]{CrisErm08}
		{Criscuoli}, S., \& {Ermolli}, I. 2008, \aap, 484, 591

	\bibitem[{{Criscuoli} \& {Rast}(2009)}]{CrisRas09}
		{Criscuoli}, S., \& {Rast}, M.~P. 2009, \aap, 495, 621

	\bibitem[{{Danilovic} {et~al.}(2010{\natexlab{a}}){Danilovic}, {Beeck},
  		{Pietarila}, {Schuessler}, {Solanki}, {Martinez Pillet}, {Bonet}, {del Toro
  		Iniesta}, {Domingo}, {Barthol}, {Berkefeld}, {Gandorfer}, {Knoelker},
  		{Schmidt}, \& {Title}}]{Dan10b}
		{Danilovic}, S., {Beeck}, B., {Pietarila}, A., {Schuessler}, M., {Solanki},
  		S.~K., {Martinez Pillet}, V., {Bonet}, J.~A., {del Toro Iniesta}, J.~C.,
  		{Domingo}, V., {Barthol}, P., {Berkefeld}, T., {Gandorfer}, A., {Knoelker},
  		M., {Schmidt}, W., \& {Title}, A.~M. 2010{\natexlab{a}}, ArXiv e-prints

	\bibitem[{{Danilovic} {et~al.}(2010{\natexlab{b}}){Danilovic}, {Sch{\"u}ssler},
  		\& {Solanki}}]{Dan10}
		{Danilovic}, S., {Sch{\"u}ssler}, M., \& {Solanki}, S.~K. 2010{\natexlab{b}},
  		\aap, 509, A76+

	\bibitem[{{de Wijn} {et~al.}(2005){de Wijn}, {Rutten}, {Haverkamp}, \&
  		{S{\"u}tterlin}}]{deW05}
		{de Wijn}, A.~G., {Rutten}, R.~J., {Haverkamp}, E.~M.~W.~P., \&
  		{S{\"u}tterlin}, P. 2005, \aap, 441, 1183
  
	\bibitem[{{de Wijn} {et~al.}(2008){de Wijn}, {Lites}, {Berger}, {Frank},
  		{Tarbell}, \& {Ishikawa}}]{deW08}
		{de Wijn}, A.~G., {Lites}, B.~W., {Berger}, T.~E., {Frank}, Z.~A., {Tarbell},
  		T.~D., \& {Ishikawa}, R. 2008, \apj, 684, 1469

	\bibitem[{{de Wijn} {et~al.}(2009){de Wijn}, {Stenflo}, {Solanki}, \&
  		{Tsuneta}}]{deW09}
		{de Wijn}, A.~G., {Stenflo}, J.~O., {Solanki}, S.~K., \& {Tsuneta}, S. 2009,
 		Space Science Reviews, 144, 275

	\bibitem[{{Deinzer} {et~al.}(1984){Deinzer}, {Hensler}, {Schussler}, \&
  		{Weisshaar}}]{Dein1984}
		{Deinzer}, W., {Hensler}, G., {Schussler}, M., \& {Weisshaar}, E. 1984, \aap,
  		139, 435

	\bibitem[{{Dunn} \& {Zirker}(1973)}]{DunZir73}
		{Dunn}, R.~B., \& {Zirker}, J.~B. 1973, \solphys, 33, 281

	\bibitem[{{Frazier} \& {Stenflo}(1972)}]{Fraz1972}
		{Frazier}, E.~N., \& {Stenflo}, J.~O. 1972, \solphys, 27, 330

	\bibitem[{{Gingerich} {et~al.}(1971){Gingerich}, {Noyes}, {Kalkofen}, \&
  		{Cuny}}]{Gin71}
		{Gingerich}, O., {Noyes}, R.~W., {Kalkofen}, W., \& {Cuny}, Y. 1971, \solphys,
 		18, 347

	\bibitem[{{Goode} {et~al.}(2010){Goode}, {Yurchyshyn}, {Cao}, {Abramenko},
  		{Andic}, {Ahn}, \& {Chae}}]{Goo10}
		{Goode}, P.~R., {Yurchyshyn}, V., {Cao}, W., {Abramenko}, V., {Andic}, A.,
		{Ahn}, K., \& {Chae}, J. 2010, \apjl, 714, L31

	\bibitem[{{Hirzberger} \& {Wiehr}(2005)}]{Hirz05}
		{Hirzberger}, J., \& {Wiehr}, E. 2005, \aap, 438, 1059

	\bibitem[{{Ishikawa} {et~al.}(2007){Ishikawa}, {Tsuneta}, {Kitakoshi},
  		{Katsukawa}, {Bonet}, {Vargas Dom{\'{\i}}nguez}, {Rouppe van der Voort},
  		{Sakamoto}, \& {Ebisuzaki}}]{Ish07}
		{Ishikawa}, R., {Tsuneta}, S., {Kitakoshi}, Y., {Katsukawa}, Y., {Bonet},
  		J.~A., {Vargas Dom{\'{\i}}nguez}, S., {Rouppe van der Voort}, L.~H.~M.,
  		{Sakamoto}, Y., \& {Ebisuzaki}, T. 2007, \aap, 472, 911

	\bibitem[{{Judge} {et~al.}(2010){Judge}, {Tritschler}, {Uitenbroek}, {Reardon},
  		{Cauzzi}, \& {de Wijn}}]{Jud10}
		{Judge}, P.~G., {Tritschler}, A., {Uitenbroek}, H., {Reardon}, K., {Cauzzi},
  		G., \& {de Wijn}, A. 2010, \apj, 710, 1486

	\bibitem[{{Keller}(1992)}]{Kel92}
		{Keller}, C.~U. 1992, \nat, 359, 307

	\bibitem[{{Knoelker} {et~al.}(1988){Knoelker}, {Schuessler}, \&
  		{Weisshaar}}]{Knol88}
		{Knoelker}, M., {Schuessler}, M., \& {Weisshaar}, E. 1988, \aap, 194, 257

	\bibitem[{{Kobel} {et~al.}(2009){Kobel}, {Hirzberger}, {Solanki}, {Gandorfer},
  		\& {Zakharov}}]{Kob09}
		{Kobel}, P., {Hirzberger}, J., {Solanki}, S.~K., {Gandorfer}, A., \&
  		{Zakharov}, V. 2009, \aap, 502, 303

	\bibitem[{{Langangen} {et~al.}(2007){Langangen}, {Carlsson}, {Rouppe van der
  		Voort}, \& {Stein}}]{Lang07}
		{Langangen}, {\O}., {Carlsson}, M., {Rouppe van der Voort}, L., \& {Stein},
  		R.~F. 2007, \apj, 655, 615

	\bibitem[{{Langhans} {et~al.}(2004){Langhans}, {Schmidt}, \&
  		{Rimmele}}]{Lan2004}
		{Langhans}, K., {Schmidt}, W., \& {Rimmele}, T. 2004, \aap, 423, 1147

	\bibitem[{{L\"{o}fdahl} {et~al.}(1998){L\"{o}fdahl}, {Berger}, {Shine}, \&
  		{Title}}]{Lof98}
		{L\"{o}fdahl}, M.~G., {Berger}, T.~E., {Shine}, R.~S., \& {Title}, A.~M. 1998,
  		\apj, 495, 965

	\bibitem[{{L\"{o}fdahl}(2002)}]{Lof02}
		{L\"{o}fdahl}, M.~G. 2002, ArXiv Physics e-prints

	\bibitem[{{Mehltretter}(1974)}]{Meh74}
		{Mehltretter}, J.~P. 1974, \solphys, 38, 43

	\bibitem[{{Muller} \& {Roudier}(1984)}]{MulRoud1984SoPh}
		{Muller}, R., \& {Roudier}, T. 1984, \solphys, 94, 33

	\bibitem[{{Muller} {et~al.}(2000){Muller}, {Dollfus}, {Montagne}, {Moity}, \&
  		{Vigneau}}]{Mul00}
		{Muller}, R., {Dollfus}, A., {Montagne}, M., {Moity}, J., \& {Vigneau}, J.
  		2000, \aap, 359, 373

	\bibitem[{{Narayan} \& {Scharmer}(2010)}]{Nar10}
		{Narayan}, G., \& {Scharmer}, G.~B. 2010, ArXiv e-prints

	\bibitem[{{Nisenson} {et~al.}(2003){Nisenson}, {van Ballegooijen}, {de Wijn},
  		\& {S{\"u}tterlin}}]{Nis03}
		{Nisenson}, P., {van Ballegooijen}, A.~A., {de Wijn}, A.~G., \&
  		{S{\"u}tterlin}, P. 2003, \apj, 587, 458

	\bibitem[{{Okunev} \& {Kneer}(2005)}]{OkKn05}
		{Okunev}, O.~V., \& {Kneer}, F. 2005, \aap, 439, 323

	\bibitem[{{Pizzo} {et~al.}(1993){Pizzo}, {MacGregor}, \& {Kunasz}}]{Pizzo1993}
		{Pizzo}, V.~J., {MacGregor}, K.~B., \& {Kunasz}, P.~B. 1993, \apj, 413, 764

	\bibitem[{{Rabin}(1992)}]{Rabin1992}
		{Rabin}, D. 1992, \solphys, 391, 832

	\bibitem[{{Reardon} \& {Cavallini}(2008)}]{ReaCav08}
		{Reardon}, K.~P., \& {Cavallini}, F. 2008, \aap, 481, 897

	\bibitem[{{Rees} \& {Semel}(1979)}]{ReeSem79}
		{Rees}, D.~E., \& {Semel}, M.~D. 1979, \aap, 74, 1

	\bibitem[{{Rimmele}(2004)}]{Rim04}
		{Rimmele}, T.~R. 2004, in Presented at the Society of Photo-Optical
  		Instrumentation Engineers (SPIE) Conference, Vol. 5490, Society of
  		Photo-Optical Instrumentation Engineers (SPIE) Conference Series, ed.
  		D.~{Bonaccini Calia}, B.~L. {Ellerbroek}, \& R.~{Ragazzoni}, 34--46

	\bibitem[{{Rouppe van der Voort} {et~al.}(2005){Rouppe van der Voort},
  		{Hansteen}, {Carlsson}, {Fossum}, {Marthinussen}, {van Noort}, \&
  		{Berger}}]{Roup05}
		{Rouppe van der Voort}, L.~H.~M., {Hansteen}, V.~H., {Carlsson}, M., {Fossum},
  		A., {Marthinussen}, E., {van Noort}, M.~J., \& {Berger}, T.~E. 2005, \aap,
  		435, 327

	\bibitem[{{Ruiz Cobo} \& {del Toro Iniesta}(1992)}]{RuiCTorI92}
		{Ruiz Cobo}, B., \& {del Toro Iniesta}, J.~C. 1992, \apj, 398, 375

	\bibitem[{{S{\'a}nchez Almeida} {et~al.}(2004){S{\'a}nchez Almeida},
  		{M{\'a}rquez}, {Bonet}, {Dom{\'{\i}}nguez Cerde{\~n}a}, \& {Muller}}]{SanA04}
		{S{\'a}nchez Almeida}, J., {M{\'a}rquez}, I., {Bonet}, J.~A., {Dom{\'{\i}}nguez
  		Cerde{\~n}a}, I., \& {Muller}, R. 2004, \apjl, 609, L91

	\bibitem[{{S{\'a}nchez Almeida} {et~al.}(2007){S{\'a}nchez Almeida}, {Teriaca},
  		{S{\"u}tterlin}, {Spadaro}, {Sch{\"u}hle}, \& {Rutten}}]{SanA07}
		{S{\'a}nchez Almeida}, J., {Teriaca}, L., {S{\"u}tterlin}, P., {Spadaro}, D.,
  		{Sch{\"u}hle}, U., \& {Rutten}, R.~J. 2007, \aap, 475, 1101
  
	\bibitem[{{S{\'a}nchez Almeida} {et~al.}(2010){S{\'a}nchez Almeida}, {Bonet},
  		{Viticchi{\'e}}, \& {Del Moro}}]{SanA10}
		{S{\'a}nchez Almeida}, J., {Bonet}, J.~A., {Viticchi{\'e}}, B., \& {Del Moro},
  		D. 2010, \apjl, 715, L26

	\bibitem[{{Sch{\"u}ssler} {et~al.}(2003){Sch{\"u}ssler}, {Shelyag},
  		{Berdyugina}, {V{\"o}gler}, \& {Solanki}}]{Sch03}
		{Sch{\"u}ssler}, M., {Shelyag}, S., {Berdyugina}, S., {V{\"o}gler}, A., \&
  		{Solanki}, S.~K. 2003, \apjl, 597, L173

	\bibitem[{{Shelyag} {et~al.}(2004){Shelyag}, {Sch{\"u}ssler}, {Solanki},
  		{Berdyugina}, \& {V{\"o}gler}}]{She04}
		{Shelyag}, S., {Sch{\"u}ssler}, M., {Solanki}, S.~K., {Berdyugina}, S.~V., \&
  		{V{\"o}gler}, A. 2004, \aap, 427, 335

	\bibitem[{{Shelyag} {et~al.}(2007){Shelyag}, {Sch{\"u}ssler}, {Solanki}, \&
  		{V{\"o}gler}}]{She07}
		{Shelyag}, S., {Sch{\"u}ssler}, M., {Solanki}, S.~K., \& {V{\"o}gler}, A. 2007,
  		\aap, 469, 731

	\bibitem[{{Spruit}(1976)}]{Spr1976}
		{Spruit}, H.~C. 1976, \solphys, 50, 269

	\bibitem[{{Steiner} {et~al.}(2001){Steiner}, {Hauschildt}, \& {Bruls}}]{Ste01}
		{Steiner}, O., {Hauschildt}, P.~H., \& {Bruls}, J. 2001, \aap, 372, L13

	\bibitem[{{Steiner}(2005)}]{Ste05}
		{Steiner}, O. 2005, \aap, 430, 691

	\bibitem[{{S{\"u}tterlin} {et~al.}(1999){S{\"u}tterlin}, {Wiehr}, \&
		{Stellmacher}}]{Sut99}
		{S{\"u}tterlin}, P., {Wiehr}, E., \& {Stellmacher}, G. 1999, \solphys, 189, 57

	\bibitem[{{Tritschler} \& {Uitenbroek}(2006)}]{Trit2006}
		{Tritschler}, A., \& {Uitenbroek}, H. 2006, \apj, 648, 741

	\bibitem[{{Utz} {et~al.}(2009{\natexlab{a}}){Utz}, {Hanslmeier}, {M{\"o}stl},
  		{Muller}, {Veronig}, \& {Muthsam}}]{Utz09}
		{Utz}, D., {Hanslmeier}, A., {M{\"o}stl}, C., {Muller}, R., {Veronig}, A., \&
  		{Muthsam}, H. 2009{\natexlab{a}}, \aap, 498, 289

	\bibitem[{{Utz} {et~al.}(2009{\natexlab{b}}){Utz}, {Hanslmeier}, {Muller},
  		{Veronig}, {Ryb{\'a}k}, \& {Muthsam}}]{Utz09b}
		{Utz}, D., {Hanslmeier}, A., {Muller}, R., {Veronig}, A., {Ryb{\'a}k}, J., \&
  		{Muthsam}, H. 2009{\natexlab{b}}, ArXiv e-prints

	\bibitem[{{van Noort} {et~al.}(2005){van Noort}, {Rouppe van der Voort}, \&
  		{L{\"o}fdahl}}]{VanN05}
		{van Noort}, M., {Rouppe van der Voort}, L., \& {L{\"o}fdahl}, M.~G. 2005,
  		\solphys, 228, 191

	\bibitem[{{Viticchi{\'e}} {et~al.}(2009){Viticchi{\'e}}, {Del Moro},
  		{Berrilli}, {Bellot Rubio}, \& {Tritschler}}]{Vit09}
		{Viticchi{\'e}}, B., {Del Moro}, D., {Berrilli}, F., {Bellot Rubio}, L., \&
  		{Tritschler}, A. 2009, \apjl, 700, L145

	\bibitem[{{Viticchi{\'e}} {et~al.}(2010){Viticchi{\'e}}, {S\'anchez Almeida},
  		{Del Moro}, \& {Berrilli}}]{Vit10}
		{Viticchi{\'e}}, B., {S\'anchez Almeida}, J., {Del Moro}, D., \& {Berrilli}.
  		2010, \aap, submitted

	\bibitem[{{V{\"o}gler} {et~al.}(2005){V{\"o}gler}, {Shelyag}, {Sch{\"u}ssler},
  		{Cattaneo}, {Emonet}, \& {Linde}}]{Vog05}
		{V{\"o}gler}, A., {Shelyag}, S., {Sch{\"u}ssler}, M., {Cattaneo}, F., {Emonet},
  		T., \& {Linde}, T. 2005, \aap, 429, 335

	\bibitem[{{Wiehr} {et~al.}(2004){Wiehr}, {Bovelet}, \& {Hirzberger}}]{Wie04}
		{Wiehr}, E., {Bovelet}, B., \& {Hirzberger}, J. 2004, \aap, 422, L63

\end{thebibliography}
	%\bibliographystyle{apj}

\end{document}